\documentclass[english, reprint, amssymb, amsmath, aps, prb, groupedaddress]{revtex4-1}
\usepackage[T1]{fontenc}
\usepackage[utf8]{inputenc}
\usepackage{color}
\usepackage{babel}
\usepackage{amstext}
\usepackage{amssymb}
\usepackage{graphicx}
\usepackage[unicode=true]{hyperref}

\begin{document}

\title{Evidence for Charging Effects in CdTe/CdMgTe Quantum Point Contacts}

\author{M.~Czapkiewicz}
\email{adgam@ifpan.edu.pl}

\author{V.~Kolkovsky}
\author{P.~Nowicki}
\author{M.~Wiater}
\author{T.~Wojciechowski}
\author{T.~Wojtowicz}
\author{J.~Wróbel}

\affiliation{Institute of Physics, Polish Academy of Sciences, al Lotników 32/46,
02-668 Warszawa, Poland}


\begin{abstract}
Here we report on fabrication and low temperature magnetotransport
measurements of quantum point contacts patterned from a novel two-dimensional
electron system --- CdTe/CdMgTe modulation doped heterostructure.
From the temperature and bias dependence we ascribe the reported data
to evidence for a weakly bound state which is naturally formed inside
a CdTe quantum constriction due to charging effects. We argue that
the spontaneous introduction of an open dot is responsible for
replacement of flat conductance plateaus by quasi-periodic resonances
with amplitude less than $2e^{2}/h$, as found in our system. Additionally,
below $1$~K a pattern of weaker conductance peaks, superimposed
upon wider resonances, is also observed. 
\end{abstract}
\maketitle

Quantum point contacts (QPCs) are conventionally considered as open mesoscopic
systems and their characteristic feature, i.e. integer quantized conductance
$G$, is well-understood as a single electron effect \cite{Wees1991}.
Nevertheless, some additional non-integer anomalous resonances are also
commonly observed at low temperatures and their exact origin is under active
debate. Generally they are attributed to electron-electron (\emph{e-e})
interactions and more specifically, to the formation of quasi-bound states inside
the constriction. Such charge droplets may reveal the Kondo physics or become,
via the exchange energy term, ferromagnetically polarized \cite{*[{}] [{, and
references therein.}] Berggren2010}. To date, the physical mechanism of
localization is still unclear and the role of charging effects
\cite{Sablikov2000} in the spontaneous formation of an open dot remains
controversial. Therefore, a semiconducting material with a large ratio of
Coulomb to kinetic energies is expedient to resolve those important issues and
is potentially useful for spintronic applications.

Motivated by these considerations we report on fabrication and low
temperature magnetotransport measurements of quantum point contacts,
patterned from \emph{n-}type CdTe/CdMgTe modulation doped quantum
well \cite{Piot2010}, not studied before in the ballistic transport
regime. It is expected that the correlations effects in CdTe are stronger,
as compared to GaAs, since the effective mass is larger and the dielectric
constant is smaller. As a result, the average distance between electrons
$r_{s}$, expressed in effective Bohr radius units, is $2.2$ times
larger for CdTe than for GaAs with the same carrier density. In this
report we provide the evidence for spontaneous formation of quasi-bound
state in short and nominally symmetric QPCs, which suggests that the
counter-intuitive appearance of localisation is caused by \emph{e-e}
correlations. This is supported by the temperature and bias dependence
of observed conductance resonances. 

\begin{figure}
\noindent \begin{centering}
\includegraphics{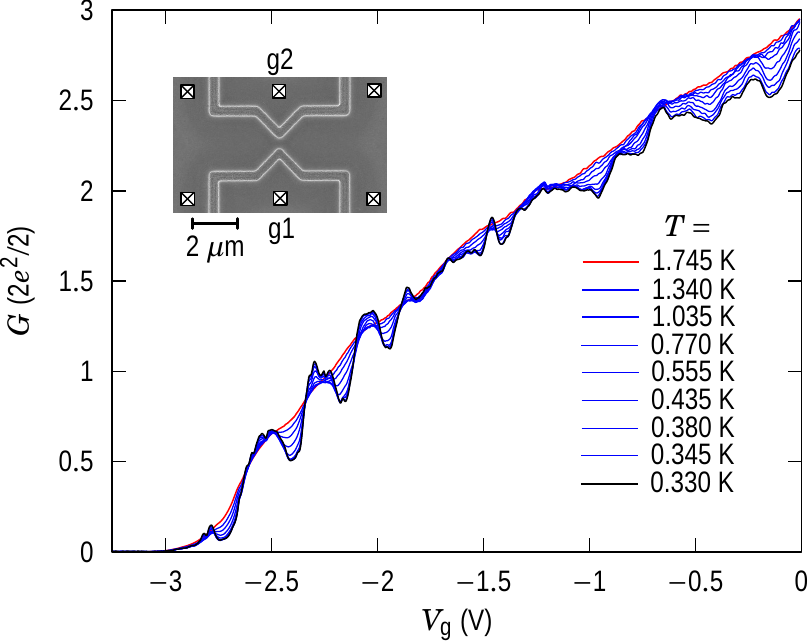}
\par\end{centering}

\caption{(color online). The zero-bias differential conductance $G=dI/dV$
(in $G_{0}=2e^{2}/h$ units) as a function of gate voltage $V_{\text{g}}$
for temperatures $T$ from $0.330$~K to $1.745$~K. The inset displays
a electron micrograph of a typical device, electrical contacts are
shown schematically, side gates are marked with $g1$ and $g2$. $V_{\text{g}}$
have been applied symmetrically, i.e. $V_{\text{g }}=V_{\text{g1}}=V_{\text{g2}}$.\label{fig:G-Vg}}
\end{figure}

Four-terminal quantum point contacts have been made of high quality
two-dimensional electron gas (2DEG) with concentration $n_{2D}=5.6\times10^{11}\text{ cm}^{-2}$
and mobility $\mu=2.3\times10^{5}\text{ cm}^{2}/\text{Vs}$. The nanojunctions
of length $L\approx0.2$~$\mu$m and lithographic width $W_{\mathrm{lith}}=0.45\pm0.01$~$\mu$m
are patterned by \emph{e}-beam lithography and deep-etching techniques.
Carrier density in those devices is controlled by means of\textsf{
V}-shaped \emph{side gates} which are separated from the constriction
area by narrow etched grooves, see inset to Fig.~\ref{fig:G-Vg}.
A modulation doped quantum well is located $70$~nm below the surface
of the device and ohmic contacts have been prepared by direct indium
soldering. The differential conductance $G$ has been measured in
He-3 cryostat as a function of \emph{dc} source-drain bias and \emph{in-plane}
magnetic field by employing a standard low-frequency lock-in technique,
with \emph{ac} excitation voltage of $50$~$\mu$V. Data collected
at different cooldowns from room temperature demonstrate similar characteristics.

We have studied several devices of the same geometry, however, perfect
conductance quantization with $G_{0}=2e^{2}/h$ plateaus is \emph{never
}observed. The electron mean free path of the 2DEG is $\ell\approx3$~$\mu$m
and chemically etched grooves are rather smooth. Also, the estimated
physical width of the constriction $W$ is much smaller than $W_{\mathrm{lith}}$
(see below) and the quantum lifetime $\tau_{\mathrm q} \approx 0.6$ ps is rather
large \cite{Piot2010}, therefore the absence of flat conductance steps is quite
unexpected. Figure~\ref{fig:G-Vg} shows differential conductance
data, obtained for the most comprehensively studied QPC sample. $G$
is not perfectly quantized vs gate voltage, however, quasi-periodic
conductance oscillations with amplitude less than $G_{0}$ are clearly
visible. Additionally, at low temperatures ($T<1.4$~K) a weaker
pattern of less regular and narrower fluctuations is observed. Some
of the random fluctuations are caused by disorder, but regular conductance
resonances with amplitude reduced down to $\approx0.6\times(2e^{2}/h)$
must be related to a subsequent population of 1D channels. The energy
distance between quantized modes $\epsilon_{1D}=2.0\pm0.5$~meV has
been obtained from the bias dependence, therefore the physical width
of our device is estimated as $W\approx40$~nm for Fermi energy $E_{F}=2.0$~meV
(assuming parabolic confinement).

\begin{figure}
\noindent \begin{centering}
\includegraphics{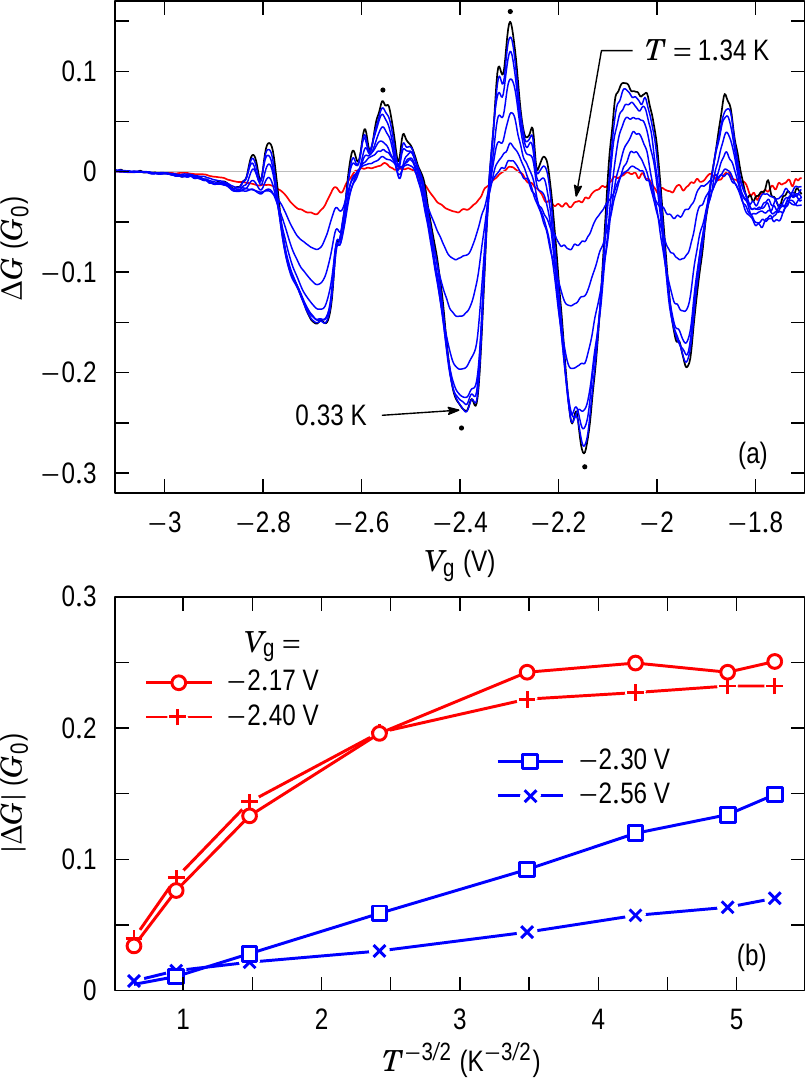}
\par\end{centering}

\caption{(color online). (a) Oscillating part of the zero-bias conductance,
defined as $\Delta G=G(T)-G(T=1.745\,\text{K})$ as a function of
temperature $T$, vs gate voltage $V_{\text{g}}$. Black line corresponds
to $T=0.33$~K, red (or gray) line corresponds to $T=1.34$~K, all
intermediate temperatures are the same as in Fig.~\ref{fig:G-Vg}.
(b) The absolute values of $\Delta G$ as a function of temperature,
plotted vs $T^{-3/2}$. Data marked with blue (dark gray) symbols
and lines have been obtained for two maxima of $\Delta G$, data marked
with red (gray) symbols and lines, for two strongest minima. Corresponding
gate voltages are indicated with small dots in (a).\label{fig:DeltaG-T}}

\end{figure}

Reduced steps and conductance resonances had been already observed
for longer quantum wires or disordered QPCs made of AlGaAs/GaAs heterostructures
\cite{Wees1991,Wrobel1992}. They are commonly attributed to a single
electron interference and back-scattering effects caused by disorder.
This mechanism is supported by the temperature dependence of conductance
traces, recorded vs $E_{\text{F}}$. At few kelvins, $G$ curves are
rather smooth and monotonic, whereas at lower $T$ fluctuations and
resonances show up and regions with positive and negative temperature
gradient appear. In an one-electron picture this is explained by thermal
averaging which smooths out an interference pattern. At first sight
a similar behaviour is observed in Fig.~\ref{fig:G-Vg}, yet a closer
look disclose important differences.

\begin{figure}
\noindent \begin{centering}
\includegraphics{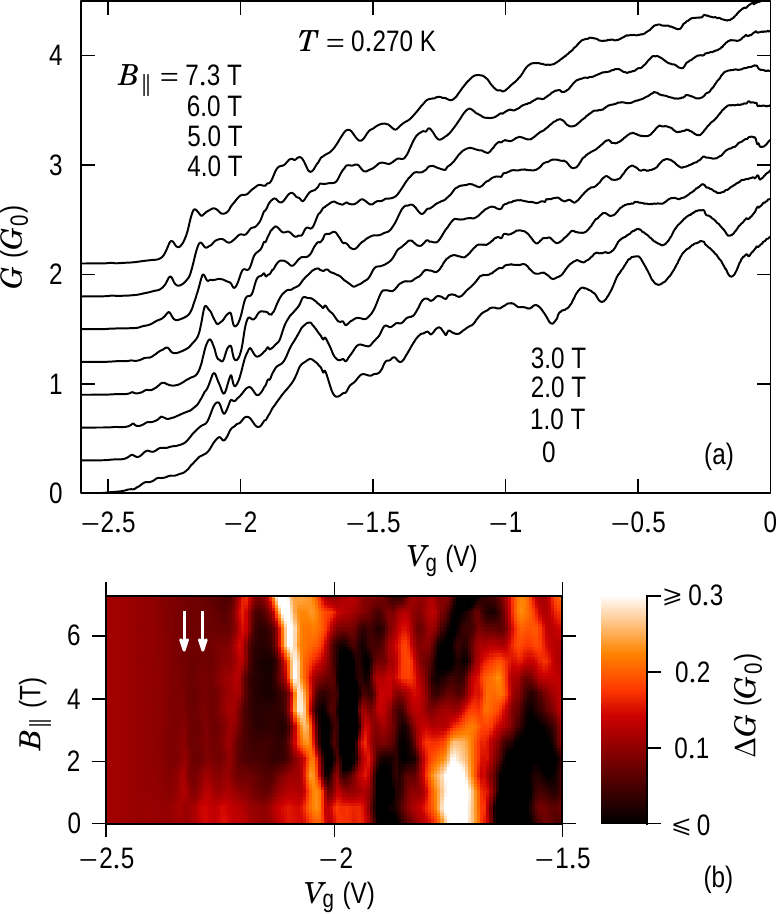}
\par\end{centering}

\caption{(color online). (a) $G(V_{\text{g}})$ for $T=0.27$~K at various
\emph{in-plain} magnetic fields $B_{\parallel}$, from $0$ (bottom)
to $7.3$~T (top), collected at different cooldown cycle than data
shown in Fig.~\ref{fig:G-Vg}. For clarity curves are shifted vertically
by offset of $0.3$. (b) Conductance $\Delta G$ oscillations as a
function of $V_{\text{g}}$ and $B_{\parallel}$, data are obtained
from $G(V_{\text{g}})$ curves by subtracting a smooth background
from each.\label{fig:G-B}}

\end{figure}

In order to show the temperature dependence of conductance resonances
in more detail, $\Delta G$ values have been calculated by subtracting
a smooth curve measured for $T=1.75$~K from all other data, collected
at lower temperatures. Results are summarized in Figure~\ref{fig:DeltaG-T}.
$\Delta G$ reveals five regular oscillations which are smoothed out
when temperature increases, however, quite distinct thermal averaging
scenarios are found, according to the $\Delta G$ sign. Conductance anti-resonances
($\Delta G<0)$, with positive temperature dependence, are practically
unchanged up to $T\approx0.5$~K and then start to disappear. At
the same time, resonances ($\Delta G>0)$ show stronger, approximately
like $1/T^{3/2}$, low temperature dependence, as demonstrated in
Fig.~\ref{fig:DeltaG-T}(b). Furthermore, up to $20$ smaller, quasi-periodic
conductance peaks are superimposed on wider oscillations up to $G \simeq 2$.
This was not reported in earlier mesoscopic studies of disordered QPCs. The overall
picture is therefore more complicated than the one provided by interference
of electron waves and suggests a role of the \emph{e-e} correlations.

Conductance data presented in Figs~\ref{fig:G-Vg} and \ref{fig:DeltaG-T}
demonstrate striking similarities with $G(V_{\text{g}})$ curves obtained
for long ($\sim0.8$~$\mu$m) GaAs quantum point contact, with centrally
embedded open quantum dot \cite{Liang1998,Tkachenko2001b}. For such
a specially designed device, conductance oscillations with reduced
height (from $0.5$ to $0.8$~$G_{0}$, depending on the dot size)
were observed in place of a flat plateaus. They were interpreted as
a Fabry-Perot like resonances, placed above the threshold of each
quantized channel. The amplitude of resonances must be reduced, because
potential barriers at the entrance and at the exit to the dot will
never be exactly the same \cite{Tkachenko2001b}. Additionally, the
Coulomb charging effects in an open dot manifested itself as a smaller
and narrower periodic peaks overlaying upon wider anti-resonances.
The amplitude of Coulomb peaks increased monotonically down to $T=0.05$~K.
Clearly, similar features are observed in Fig.~\ref{fig:DeltaG-T}.
Therefore, narrow resonances with $T^{-1.5}$ temperature dependence
may be related to the quantization of charge which gradually fills
shallow potential well when gate voltage increases. Recently,
residual charge quantization has been directly observed for GaAs open
quantum dot by capacitance measurements \cite{Amasha2011}.

\begin{figure}[t]
\noindent \begin{centering}
\includegraphics{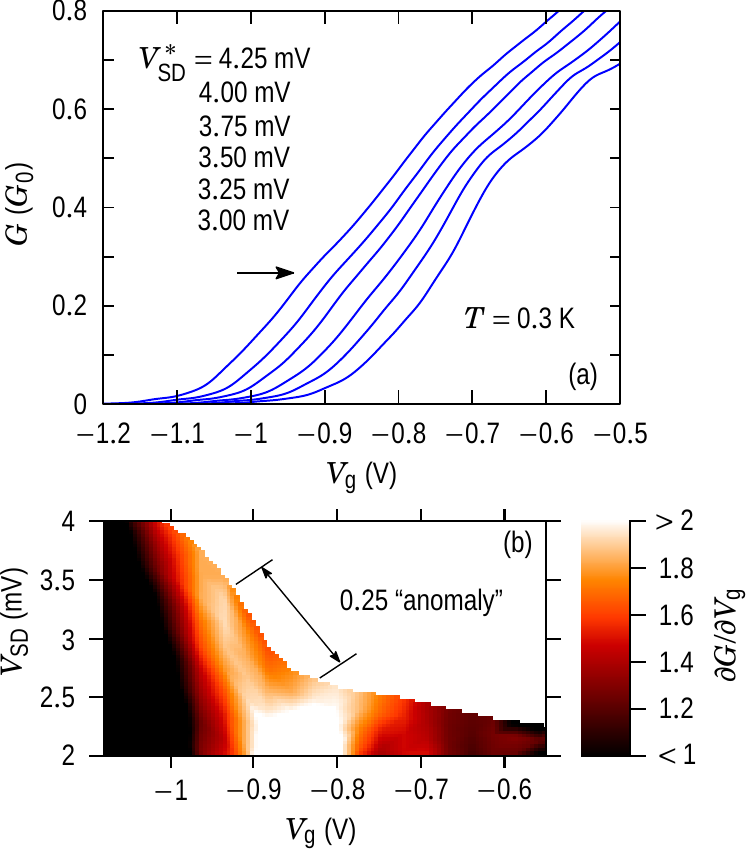}
\par\end{centering}

\caption{(color online). (a) $G(V_{\text{g}})$ for $T=0.3$~K at various
\emph{source-drain} voltages $V_{\text{SD}}^{*}$, from $4.25$ (left)
to $3.0$~mV (right). Curves are successively shifted horizontally
by $0.025$~V for clarity. (b) Transconductance $\partial G/\partial V_{\text{g}}$
as a function of $V_{\text{g}}$ and $V_{\text{SD}}$, where $V_{\text{SD}}$
is the true dc bias, obtained by subtracting a voltage drop on serial
contact resistances. Data differ with previous on pinch-off voltage,
because have been measured for a subsequent cooldown cycle. \label{fig:G-Vsd}}
\end{figure}

Self-consistent calculations suggest that quasi-localized states
exist also in a large class of smoothly varying constrictions patterned
\emph{without} the intentional central widening. When interactions
at wider regions of QPC are taken into account, the narrowest middle
part may be spontaneously charged and form an open dot \cite{Hirose2003}.
This effect is very length dependent --- for very short constrictions
bound states are not created, whereas for longer devices a chain of
charge droplets is predicted to occur \cite{Rejec2006}. If the device
is sufficiently long ($L>100$~nm, for GaAs) quasi-localized states
may develop also for higher densities, when more then one channel
is occupied \cite{Ihnatsenka2007}. 

Nevertheless, the physical mechanism of electron localization remains
unclear and its influence on conductance measurements is highly debated.
Recently it has been proposed \cite{Song2011} that longitudinal resonant levels are
formed within the constriction due to momentum mismatch and interference
effects. In contrary, previous literature suggests \cite{Shchamkhalova2007}
that the creation of quasi-bound states is caused by Friedel oscillations
of electron density which emerge at two opposite sides of the QPC
and form potential barriers which surround the central part of a device.
A barrier for the open channel arises from the interaction with electrons
from all closed channels, which are reflected at side boundaries.

Moreover, if Friedel oscillations (FO) are responsible for barrier formation,
the energy averaging must lead to an overall increase in conductance
with increased temperature, since then barriers are smoothed. This
scenario has recently been supported by numerical simulations and observed
experimentally for a clean GaAs quantum wires at $G>2$ (in $G_{0}$
units) \cite{Renard2008}. A similar effect is also visible in
Fig.~\ref{fig:G-Vg}, conductance resonances are smoothed with temperature but
their averaged value increases with $T$, specially for $G\gtrsim1.5$. We expect
that the influence of FO on transport properties will be larger in
CdTe than GaAs, however, such temperature dependence may be also caused
by a weak localization effect from the wider regions of our sample.

If weakly bound states are formed, calculations show that the exchange
interaction creates a spin-imbalance below the first conductance plateau
\cite{Rejec2006}. This prediction is supported by a large number of experiments
carried out for GaAs QPCs \cite{*[{See, e.g., the review }] [{, and references
therein.}] Micolich2011}. Figure \ref{fig:G-B} shows the effect of an in-plane
magnetic field on our device. The spin-splitting of conductance resonances
becomes clearly visible. It is evident from Fig.~\ref{fig:G-B}(b), that already
at $B_{\parallel}\approx7$~T the energy of spin-up level of the lowest mode
crosses the energy of spin-down level of the higher channel. However, a magnetic
field that is sufficient to separate the spins, does not change the shape and
location of weaker conductance resonances which are observed near pinch-off
(marked with arrows). The same key feature was observed also for a
quasi-ballistic GaAs quantum wire in the vicinity of a so-called ``0.7-anomaly''
\cite{Czapkiewicz2008}. 

We thus find a magnetized regime below the first
conduction resonance, in agreement with a picture of spin-polarized charge
droplets formed within the QPC \cite{Ihnatsenka2007}. On the other hand, for
$G<1$ (in $G_{0}$ units) the spin degeneracy of transport channels may be
spontaneously lifted without the appearance of localized states
\cite{Jaksch2006,Lassl2007}. By combining bias dependence with magnetic field
data shown in Fig.~\ref{fig:G-B} we have been able to determine the effective
Land\'e factor $g^{*}$ at $G>1$. We have found that $|g^{*}|=1.7\pm0.1$ in
remarkable agreement with bulk value reported for CdTe \cite{Oestreich1996}.

An additional argument for natural formation of an open dot in our
device is provided by the conductance data measured for large source-drain
voltages ($V_{\text{SD}}>\epsilon_{1D}$). Calculations suggest \cite{Ihnatsenka2007b},
that if a weakly bound state is present in the constriction, its local
density of states follows Fermi energy when gate voltage is changed.
As a result, when electrons travel only in one direction (at finite
biases), the pinning of the resonant level leads to the appearance
of a plateau-like feature at $G\approx0.25$ on $G(V_{\text{g}})$
curves \cite{Ihnatsenka2009}. Such ``$0.25$-anomaly'' was measured
for rather long (\,$L=0.4$ to $1.0$~$\mu$m) GaAs quantum wires\emph{
\cite{Cronenwett2002,Sfigakis2008,Czapkiewicz2008},} where the weakly
bound states were present or were probably induced by strong
bias\cite{Sablikov2000,Ihnatsenka2009}. A similar feature is observed
also in our shorter device, as it follows from Fig.~\ref{fig:G-Vsd}.
The formation of a weak shoulder at $G\approx0.3$ with increasing
source-drain voltage is marked with an arrow. The shift of the conduction
onset towards lower values of $V_{\text{g}}$ and the appearance of
``$0.25$-anomaly'' is more clearly visible at the transconductance
$\partial G/\partial V_{\text{g}}$ color map, displayed in Fig.~\ref{fig:G-Vsd}(b).

In summary, we have presented low-temperature conductance measurements
on short point contact made of a CdTe/CdMgTe quantum well. Many of the observed
features are shared by longer GaAs wires with centrally embedded open
quantum dots. This suggests the spontaneous formation of potential barriers at
the entrance and at the exit of our device. Recently, the disappearance of quantized
plateaus and evidence for the formation of quasi-bound states in an asymmetrical
GaAs QPC have been also reported \cite{Wu2012} and attributed to an abrupt rise
of confining potential along the channel. Our sample is symmetric and constriction is defined by
an adiabatic variation of width, so a momentum mismatch is less severe, yet
stronger \emph{e-e} interactions in CdTe may induce a natural bound state
considerably easier. This makes cadmium telluride a promising host
material for studying the interplay between interference and correlation
effects in low dimensions.

We acknowledge the support from the Polish Ministry of Science and
Higher Education, project number N202/103936 and partial support by
the European Union within European Regional Development Fund (grant
Innovative Economy POIG.01.01.02-00-008/08).


\begin{thebibliography}{24}%
\makeatletter
\providecommand \@ifxundefined [1]{%
 \@ifx{#1\undefined}
}%
\providecommand \@ifnum [1]{%
 \ifnum #1\expandafter \@firstoftwo
 \else \expandafter \@secondoftwo
 \fi
}%
\providecommand \@ifx [1]{%
 \ifx #1\expandafter \@firstoftwo
 \else \expandafter \@secondoftwo
 \fi
}%
\providecommand \natexlab [1]{#1}%
\providecommand \enquote  [1]{``#1''}%
\providecommand \bibnamefont  [1]{#1}%
\providecommand \bibfnamefont [1]{#1}%
\providecommand \citenamefont [1]{#1}%
\providecommand \href@noop [0]{\@secondoftwo}%
\providecommand \href [0]{\begingroup \@sanitize@url \@href}%
\providecommand \@href[1]{\@@startlink{#1}\@@href}%
\providecommand \@@href[1]{\endgroup#1\@@endlink}%
\providecommand \@sanitize@url [0]{\catcode `\\12\catcode `\$12\catcode
  `\&12\catcode `\#12\catcode `\^12\catcode `\_12\catcode `\%12\relax}%
\providecommand \@@startlink[1]{}%
\providecommand \@@endlink[0]{}%
\providecommand \url  [0]{\begingroup\@sanitize@url \@url }%
\providecommand \@url [1]{\endgroup\@href {#1}{\urlprefix }}%
\providecommand \urlprefix  [0]{URL }%
\providecommand \Eprint [0]{\href }%
\providecommand \doibase [0]{http://dx.doi.org/}%
\providecommand \selectlanguage [0]{\@gobble}%
\providecommand \bibinfo  [0]{\@secondoftwo}%
\providecommand \bibfield  [0]{\@secondoftwo}%
\providecommand \translation [1]{[#1]}%
\providecommand \BibitemOpen [0]{}%
\providecommand \bibitemStop [0]{}%
\providecommand \bibitemNoStop [0]{.\EOS\space}%
\providecommand \EOS [0]{\spacefactor3000\relax}%
\providecommand \BibitemShut  [1]{\csname bibitem#1\endcsname}%
\let\auto@bib@innerbib\@empty
\bibitem [{\citenamefont {van Wees}\ \emph {et~al.}(1991)\citenamefont {van
  Wees}, \citenamefont {Kouwenhoven}, \citenamefont {Willems}, \citenamefont
  {Harmans}, \citenamefont {Mooij}, \citenamefont {van Houten}, \citenamefont
  {Beenakker}, \citenamefont {Williamson},\ and\ \citenamefont
  {Foxon}}]{Wees1991}%
  \BibitemOpen
  \bibfield  {author} {\bibinfo {author} {\bibfnamefont {B.~J.}\ \bibnamefont
  {van Wees}}, \bibinfo {author} {\bibfnamefont {L.~P.}\ \bibnamefont
  {Kouwenhoven}}, \bibinfo {author} {\bibfnamefont {E.~M.~M.}\ \bibnamefont
  {Willems}}, \bibinfo {author} {\bibfnamefont {C.~J. P.~M.}\ \bibnamefont
  {Harmans}}, \bibinfo {author} {\bibfnamefont {J.~E.}\ \bibnamefont {Mooij}},
  \bibinfo {author} {\bibfnamefont {H.}~\bibnamefont {van Houten}}, \bibinfo
  {author} {\bibfnamefont {C.~W.~J.}\ \bibnamefont {Beenakker}}, \bibinfo
  {author} {\bibfnamefont {J.~G.}\ \bibnamefont {Williamson}}, \ and\ \bibinfo
  {author} {\bibfnamefont {C.~T.}\ \bibnamefont {Foxon}},\ }\href
  {http://link.aps.org/doi/10.1103/PhysRevB.43.12431} {\bibfield  {journal}
  {\bibinfo  {journal} {Phys. Rev. B}\ }\textbf {\bibinfo {volume} {43}},\
  \bibinfo {pages} {12431} (\bibinfo {year} {1991})}\BibitemShut {NoStop}%
\bibitem [{\citenamefont {Berggren}\ and\ \citenamefont
  {Pepper}(2010)}]{Berggren2010}%
  \BibitemOpen
  \bibfield  {author} {\bibinfo {author} {\bibfnamefont {K.-F.}\ \bibnamefont
  {Berggren}}\ and\ \bibinfo {author} {\bibfnamefont {M.}~\bibnamefont
  {Pepper}},\ }\href {\doibase 10.1098/rsta.2009.0226} {\bibfield  {journal}
  {\bibinfo  {journal} {Phil. Trans. R. Soc. A}\ }\textbf {\bibinfo {volume}
  {368}},\ \bibinfo {pages} {1141} (\bibinfo {year} {2010})}\BibitemShut
  {NoStop}%
\bibitem [{\citenamefont {Sablikov}\ \emph {et~al.}(2000)\citenamefont
  {Sablikov}, \citenamefont {Polyakov},\ and\ \citenamefont
  {Büttiker}}]{Sablikov2000}%
  \BibitemOpen
  \bibfield  {author} {\bibinfo {author} {\bibfnamefont {V.~A.}\ \bibnamefont
  {Sablikov}}, \bibinfo {author} {\bibfnamefont {S.~V.}\ \bibnamefont
  {Polyakov}}, \ and\ \bibinfo {author} {\bibfnamefont {M.}~\bibnamefont
  {Büttiker}},\ }\href {http://link.aps.org/doi/10.1103/PhysRevB.61.13763}
  {\bibfield  {journal} {\bibinfo  {journal} {Phys. Rev. B}\ }\textbf {\bibinfo
  {volume} {61}},\ \bibinfo {pages} {13763} (\bibinfo {year}
  {2000})}\BibitemShut {NoStop}%
\bibitem [{\citenamefont {Piot}\ \emph {et~al.}(2010)\citenamefont {Piot},
  \citenamefont {Kunc}, \citenamefont {Potemski}, \citenamefont {Maude},
  \citenamefont {Betthausen}, \citenamefont {Vogl}, \citenamefont {Weiss},
  \citenamefont {Karczewski},\ and\ \citenamefont {Wojtowicz}}]{Piot2010}%
  \BibitemOpen
  \bibfield  {author} {\bibinfo {author} {\bibfnamefont {B.~A.}\ \bibnamefont
  {Piot}}, \bibinfo {author} {\bibfnamefont {J.}~\bibnamefont {Kunc}}, \bibinfo
  {author} {\bibfnamefont {M.}~\bibnamefont {Potemski}}, \bibinfo {author}
  {\bibfnamefont {D.~K.}\ \bibnamefont {Maude}}, \bibinfo {author}
  {\bibfnamefont {C.}~\bibnamefont {Betthausen}}, \bibinfo {author}
  {\bibfnamefont {A.}~\bibnamefont {Vogl}}, \bibinfo {author} {\bibfnamefont
  {D.}~\bibnamefont {Weiss}}, \bibinfo {author} {\bibfnamefont
  {G.}~\bibnamefont {Karczewski}}, \ and\ \bibinfo {author} {\bibfnamefont
  {T.}~\bibnamefont {Wojtowicz}},\ }\href
  {http://link.aps.org/doi/10.1103/PhysRevB.82.081307} {\bibfield  {journal}
  {\bibinfo  {journal} {Phys. Rev. B}\ }\textbf {\bibinfo {volume} {82}},\
  \bibinfo {pages} {081307} (\bibinfo {year} {2010})}\BibitemShut {NoStop}%
\bibitem [{\citenamefont {Wróbel}\ \emph {et~al.}(1992)\citenamefont
  {Wróbel}, \citenamefont {Kuchar}, \citenamefont {Ismail}, \citenamefont
  {Lee}, \citenamefont {Nickel},\ and\ \citenamefont {Schlapp}}]{Wrobel1992}%
  \BibitemOpen
  \bibfield  {author} {\bibinfo {author} {\bibfnamefont {J.}~\bibnamefont
  {Wróbel}}, \bibinfo {author} {\bibfnamefont {F.}~\bibnamefont {Kuchar}},
  \bibinfo {author} {\bibfnamefont {K.}~\bibnamefont {Ismail}}, \bibinfo
  {author} {\bibfnamefont {K.}~\bibnamefont {Lee}}, \bibinfo {author}
  {\bibfnamefont {H.}~\bibnamefont {Nickel}}, \ and\ \bibinfo {author}
  {\bibfnamefont {W.}~\bibnamefont {Schlapp}},\ }\href
  {http://www.sciencedirect.com/science/article/pii/003960289290348A}
  {\bibfield  {journal} {\bibinfo  {journal} {Surface Science}\ }\textbf
  {\bibinfo {volume} {263}},\ \bibinfo {pages} {261} (\bibinfo {year}
  {1992})}\BibitemShut {NoStop}%
\bibitem [{\citenamefont {Liang}\ \emph {et~al.}(1998)\citenamefont {Liang},
  \citenamefont {Simmons}, \citenamefont {Smith}, \citenamefont {Kim},
  \citenamefont {Ritchie},\ and\ \citenamefont {Pepper}}]{Liang1998}%
  \BibitemOpen
  \bibfield  {author} {\bibinfo {author} {\bibfnamefont {C.-T.}\ \bibnamefont
  {Liang}}, \bibinfo {author} {\bibfnamefont {M.~Y.}\ \bibnamefont {Simmons}},
  \bibinfo {author} {\bibfnamefont {C.~G.}\ \bibnamefont {Smith}}, \bibinfo
  {author} {\bibfnamefont {G.~H.}\ \bibnamefont {Kim}}, \bibinfo {author}
  {\bibfnamefont {D.~A.}\ \bibnamefont {Ritchie}}, \ and\ \bibinfo {author}
  {\bibfnamefont {M.}~\bibnamefont {Pepper}},\ }\href
  {http://link.aps.org/doi/10.1103/PhysRevLett.81.3507} {\bibfield  {journal}
  {\bibinfo  {journal} {Phys. Rev. Lett.}\ }\textbf {\bibinfo {volume} {81}},\
  \bibinfo {pages} {3507} (\bibinfo {year} {1998})}\BibitemShut {NoStop}%
\bibitem [{\citenamefont {Tkachenko}\ \emph {et~al.}(2001)\citenamefont
  {Tkachenko}, \citenamefont {Tkachenko}, \citenamefont {Baksheyev},
  \citenamefont {Liang}, \citenamefont {Simmons}, \citenamefont {Smith},
  \citenamefont {Ritchie}, \citenamefont {Kim},\ and\ \citenamefont
  {Pepper}}]{Tkachenko2001b}%
  \BibitemOpen
  \bibfield  {author} {\bibinfo {author} {\bibfnamefont {O.~A.}\ \bibnamefont
  {Tkachenko}}, \bibinfo {author} {\bibfnamefont {V.~A.}\ \bibnamefont
  {Tkachenko}}, \bibinfo {author} {\bibfnamefont {D.~G.}\ \bibnamefont
  {Baksheyev}}, \bibinfo {author} {\bibfnamefont {C.-T.}\ \bibnamefont
  {Liang}}, \bibinfo {author} {\bibfnamefont {M.~Y.}\ \bibnamefont {Simmons}},
  \bibinfo {author} {\bibfnamefont {C.~G.}\ \bibnamefont {Smith}}, \bibinfo
  {author} {\bibfnamefont {D.~A.}\ \bibnamefont {Ritchie}}, \bibinfo {author}
  {\bibfnamefont {G.-H.}\ \bibnamefont {Kim}}, \ and\ \bibinfo {author}
  {\bibfnamefont {M.}~\bibnamefont {Pepper}},\ }\href
  {http://stacks.iop.org/0953-8984/13/i=42/a=312} {\bibfield  {journal}
  {\bibinfo  {journal} {J. Phys.: Condens. Matter}\ }\textbf {\bibinfo {volume}
  {13}},\ \bibinfo {pages} {9515} (\bibinfo {year} {2001})}\BibitemShut
  {NoStop}%
\bibitem [{\citenamefont {Amasha}\ \emph {et~al.}(2011)\citenamefont {Amasha},
  \citenamefont {Rau}, \citenamefont {Grobis}, \citenamefont {Potok},
  \citenamefont {Shtrikman},\ and\ \citenamefont
  {Goldhaber-Gordon}}]{Amasha2011}%
  \BibitemOpen
  \bibfield  {author} {\bibinfo {author} {\bibfnamefont {S.}~\bibnamefont
  {Amasha}}, \bibinfo {author} {\bibfnamefont {I.~G.}\ \bibnamefont {Rau}},
  \bibinfo {author} {\bibfnamefont {M.}~\bibnamefont {Grobis}}, \bibinfo
  {author} {\bibfnamefont {R.~M.}\ \bibnamefont {Potok}}, \bibinfo {author}
  {\bibfnamefont {H.}~\bibnamefont {Shtrikman}}, \ and\ \bibinfo {author}
  {\bibfnamefont {D.}~\bibnamefont {Goldhaber-Gordon}},\ }\href
  {http://link.aps.org/doi/10.1103/PhysRevLett.107.216804} {\bibfield
  {journal} {\bibinfo  {journal} {Phys. Rev. Lett.}\ }\textbf {\bibinfo
  {volume} {107}},\ \bibinfo {pages} {216804} (\bibinfo {year}
  {2011})}\BibitemShut {NoStop}%
\bibitem [{\citenamefont {Hirose}\ \emph {et~al.}(2003)\citenamefont {Hirose},
  \citenamefont {Meir},\ and\ \citenamefont {Wingreen}}]{Hirose2003}%
  \BibitemOpen
  \bibfield  {author} {\bibinfo {author} {\bibfnamefont {K.}~\bibnamefont
  {Hirose}}, \bibinfo {author} {\bibfnamefont {Y.}~\bibnamefont {Meir}}, \ and\
  \bibinfo {author} {\bibfnamefont {N.~S.}\ \bibnamefont {Wingreen}},\ }\href
  {http://link.aps.org/doi/10.1103/PhysRevLett.90.026804} {\bibfield  {journal}
  {\bibinfo  {journal} {Phys. Rev. Lett.}\ }\textbf {\bibinfo {volume} {90}},\
  \bibinfo {pages} {026804} (\bibinfo {year} {2003})}\BibitemShut {NoStop}%
\bibitem [{\citenamefont {Rejec}\ and\ \citenamefont {Meir}(2006)}]{Rejec2006}%
  \BibitemOpen
  \bibfield  {author} {\bibinfo {author} {\bibfnamefont {T.}~\bibnamefont
  {Rejec}}\ and\ \bibinfo {author} {\bibfnamefont {Y.}~\bibnamefont {Meir}},\
  }\href {\doibase 10.1038/nature05054} {\bibfield  {journal} {\bibinfo
  {journal} {Nature}\ }\textbf {\bibinfo {volume} {442}},\  (\bibinfo {year}
  {2006})}\BibitemShut {NoStop}%
\bibitem [{\citenamefont {Ihnatsenka}\ and\ \citenamefont
  {Zozoulenko}(2007)}]{Ihnatsenka2007}%
  \BibitemOpen
  \bibfield  {author} {\bibinfo {author} {\bibfnamefont {S.}~\bibnamefont
  {Ihnatsenka}}\ and\ \bibinfo {author} {\bibfnamefont {I.~V.}\ \bibnamefont
  {Zozoulenko}},\ }\href {http://link.aps.org/doi/10.1103/PhysRevB.76.045338}
  {\bibfield  {journal} {\bibinfo  {journal} {Phys. Rev. B}\ }\textbf {\bibinfo
  {volume} {76}},\ \bibinfo {pages} {045338} (\bibinfo {year}
  {2007})}\BibitemShut {NoStop}%
\bibitem [{\citenamefont {Song}\ and\ \citenamefont {Ahn}(2011)}]{Song2011}%
  \BibitemOpen
  \bibfield  {author} {\bibinfo {author} {\bibfnamefont {T.}~\bibnamefont
  {Song}}\ and\ \bibinfo {author} {\bibfnamefont {K.-H.}\ \bibnamefont {Ahn}},\
  }\href {http://link.aps.org/doi/10.1103/PhysRevLett.106.057203} {\bibfield
  {journal} {\bibinfo  {journal} {Phys. Rev. Lett.}\ }\textbf {\bibinfo
  {volume} {106}},\ \bibinfo {pages} {057203} (\bibinfo {year}
  {2011})}\BibitemShut {NoStop}%
\bibitem [{\citenamefont {Shchamkhalova}\ and\ \citenamefont
  {Sablikov}(2007)}]{Shchamkhalova2007}%
  \BibitemOpen
  \bibfield  {author} {\bibinfo {author} {\bibfnamefont {B.~S.}\ \bibnamefont
  {Shchamkhalova}}\ and\ \bibinfo {author} {\bibfnamefont {V.~A.}\ \bibnamefont
  {Sablikov}},\ }\href {http://stacks.iop.org/0953-8984/19/i=15/a=156221}
  {\bibfield  {journal} {\bibinfo  {journal} {J. of Phys.: Condens. Matter}\
  }\textbf {\bibinfo {volume} {19}},\ \bibinfo {pages} {156221} (\bibinfo
  {year} {2007})}\BibitemShut {NoStop}%
\bibitem [{\citenamefont {Renard}\ \emph {et~al.}(2008)\citenamefont {Renard},
  \citenamefont {Tkachenko}, \citenamefont {Tkachenko}, \citenamefont {Ota},
  \citenamefont {Kumada}, \citenamefont {Portal},\ and\ \citenamefont
  {Hirayama}}]{Renard2008}%
  \BibitemOpen
  \bibfield  {author} {\bibinfo {author} {\bibfnamefont {V.~T.}\ \bibnamefont
  {Renard}}, \bibinfo {author} {\bibfnamefont {O.~A.}\ \bibnamefont
  {Tkachenko}}, \bibinfo {author} {\bibfnamefont {V.~A.}\ \bibnamefont
  {Tkachenko}}, \bibinfo {author} {\bibfnamefont {T.}~\bibnamefont {Ota}},
  \bibinfo {author} {\bibfnamefont {N.}~\bibnamefont {Kumada}}, \bibinfo
  {author} {\bibfnamefont {J.-C.}\ \bibnamefont {Portal}}, \ and\ \bibinfo
  {author} {\bibfnamefont {Y.}~\bibnamefont {Hirayama}},\ }\href
  {http://link.aps.org/doi/10.1103/PhysRevLett.100.186801} {\bibfield
  {journal} {\bibinfo  {journal} {Phys. Rev. Lett.}\ }\textbf {\bibinfo
  {volume} {100}},\ \bibinfo {pages} {186801} (\bibinfo {year}
  {2008})}\BibitemShut {NoStop}%
\bibitem [{\citenamefont {Micolich}(2011)}]{Micolich2011}%
  \BibitemOpen
  \bibfield  {author} {\bibinfo {author} {\bibfnamefont {A.~P.}\ \bibnamefont
  {Micolich}},\ }\href {http://stacks.iop.org/0953-8984/23/i=44/a=443201}
  {\bibfield  {journal} {\bibinfo  {journal} {J. Phys.: Condens. Matter}\
  }\textbf {\bibinfo {volume} {23}},\ \bibinfo {pages} {443201} (\bibinfo
  {year} {2011})}\BibitemShut {NoStop}%
\bibitem [{\citenamefont {Czapkiewicz}\ \emph {et~al.}(2008)\citenamefont
  {Czapkiewicz}, \citenamefont {Zagrajek}, \citenamefont {Wróbel},
  \citenamefont {Grabecki}, \citenamefont {Fronc}, \citenamefont {Dietl},
  \citenamefont {Ohno}, \citenamefont {Matsuzaka},\ and\ \citenamefont
  {Ohno}}]{Czapkiewicz2008}%
  \BibitemOpen
  \bibfield  {author} {\bibinfo {author} {\bibfnamefont {M.}~\bibnamefont
  {Czapkiewicz}}, \bibinfo {author} {\bibfnamefont {P.}~\bibnamefont
  {Zagrajek}}, \bibinfo {author} {\bibfnamefont {J.}~\bibnamefont {Wróbel}},
  \bibinfo {author} {\bibfnamefont {G.}~\bibnamefont {Grabecki}}, \bibinfo
  {author} {\bibfnamefont {K.}~\bibnamefont {Fronc}}, \bibinfo {author}
  {\bibfnamefont {T.}~\bibnamefont {Dietl}}, \bibinfo {author} {\bibfnamefont
  {Y.}~\bibnamefont {Ohno}}, \bibinfo {author} {\bibfnamefont {S.}~\bibnamefont
  {Matsuzaka}}, \ and\ \bibinfo {author} {\bibfnamefont {H.}~\bibnamefont
  {Ohno}},\ }\href {http://stacks.iop.org/0295-5075/82/i=2/a=27003} {\bibfield
  {journal} {\bibinfo  {journal} {Europhys. Lett.}\ }\textbf {\bibinfo {volume}
  {82}},\ \bibinfo {pages} {27003} (\bibinfo {year} {2008})}\BibitemShut
  {NoStop}%
\bibitem [{\citenamefont {Jaksch}\ \emph {et~al.}(2006)\citenamefont {Jaksch},
  \citenamefont {Yakimenko},\ and\ \citenamefont {Berggren}}]{Jaksch2006}%
  \BibitemOpen
  \bibfield  {author} {\bibinfo {author} {\bibfnamefont {P.}~\bibnamefont
  {Jaksch}}, \bibinfo {author} {\bibfnamefont {I.}~\bibnamefont {Yakimenko}}, \
  and\ \bibinfo {author} {\bibfnamefont {K.-F.}\ \bibnamefont {Berggren}},\
  }\href {http://link.aps.org/doi/10.1103/PhysRevB.74.235320} {\bibfield
  {journal} {\bibinfo  {journal} {Phys. Rev. B}\ }\textbf {\bibinfo {volume}
  {74}},\ \bibinfo {pages} {235320} (\bibinfo {year} {2006})}\BibitemShut
  {NoStop}%
\bibitem [{\citenamefont {Lassl}\ \emph {et~al.}(2007)\citenamefont {Lassl},
  \citenamefont {Schlagheck},\ and\ \citenamefont {Richter}}]{Lassl2007}%
  \BibitemOpen
  \bibfield  {author} {\bibinfo {author} {\bibfnamefont {A.}~\bibnamefont
  {Lassl}}, \bibinfo {author} {\bibfnamefont {P.}~\bibnamefont {Schlagheck}}, \
  and\ \bibinfo {author} {\bibfnamefont {K.}~\bibnamefont {Richter}},\ }\href
  {http://link.aps.org/doi/10.1103/PhysRevB.75.045346} {\bibfield  {journal}
  {\bibinfo  {journal} {Phys. Rev. B}\ }\textbf {\bibinfo {volume} {75}},\
  \bibinfo {pages} {045346} (\bibinfo {year} {2007})}\BibitemShut {NoStop}%
\bibitem [{\citenamefont {Oestreich}\ \emph {et~al.}(1996)\citenamefont
  {Oestreich}, \citenamefont {Hallstein}, \citenamefont {Heberle},
  \citenamefont {Eberl}, \citenamefont {Bauser},\ and\ \citenamefont
  {R\"uhle}}]{Oestreich1996}%
  \BibitemOpen
  \bibfield  {author} {\bibinfo {author} {\bibfnamefont {M.}~\bibnamefont
  {Oestreich}}, \bibinfo {author} {\bibfnamefont {S.}~\bibnamefont
  {Hallstein}}, \bibinfo {author} {\bibfnamefont {A.~P.}\ \bibnamefont
  {Heberle}}, \bibinfo {author} {\bibfnamefont {K.}~\bibnamefont {Eberl}},
  \bibinfo {author} {\bibfnamefont {E.}~\bibnamefont {Bauser}}, \ and\ \bibinfo
  {author} {\bibfnamefont {W.~W.}\ \bibnamefont {R\"uhle}},\ }\href
  {http://link.aps.org/doi/10.1103/PhysRevB.53.7911} {\bibfield  {journal}
  {\bibinfo  {journal} {Phys. Rev. B}\ }\textbf {\bibinfo {volume} {53}},\
  \bibinfo {pages} {7911} (\bibinfo {year} {1996})}\BibitemShut {NoStop}%
\bibitem [{\citenamefont {Ihnatsenka}\ \emph {et~al.}(2007)\citenamefont
  {Ihnatsenka}, \citenamefont {Zozoulenko},\ and\ \citenamefont
  {Willander}}]{Ihnatsenka2007b}%
  \BibitemOpen
  \bibfield  {author} {\bibinfo {author} {\bibfnamefont {S.}~\bibnamefont
  {Ihnatsenka}}, \bibinfo {author} {\bibfnamefont {I.~V.}\ \bibnamefont
  {Zozoulenko}}, \ and\ \bibinfo {author} {\bibfnamefont {M.}~\bibnamefont
  {Willander}},\ }\href {http://link.aps.org/doi/10.1103/PhysRevB.75.235307}
  {\bibfield  {journal} {\bibinfo  {journal} {Phys. Rev. B}\ }\textbf {\bibinfo
  {volume} {75}},\ \bibinfo {pages} {235307} (\bibinfo {year}
  {2007})}\BibitemShut {NoStop}%
\bibitem [{\citenamefont {Ihnatsenka}\ and\ \citenamefont
  {Zozoulenko}(2009)}]{Ihnatsenka2009}%
  \BibitemOpen
  \bibfield  {author} {\bibinfo {author} {\bibfnamefont {S.}~\bibnamefont
  {Ihnatsenka}}\ and\ \bibinfo {author} {\bibfnamefont {I.~V.}\ \bibnamefont
  {Zozoulenko}},\ }\href {http://link.aps.org/doi/10.1103/PhysRevB.79.235313}
  {\bibfield  {journal} {\bibinfo  {journal} {Phys. Rev. B}\ }\textbf {\bibinfo
  {volume} {79}},\ \bibinfo {pages} {235313} (\bibinfo {year}
  {2009})}\BibitemShut {NoStop}%
\bibitem [{\citenamefont {Cronenwett}\ \emph {et~al.}(2002)\citenamefont
  {Cronenwett}, \citenamefont {Lynch}, \citenamefont {Goldhaber-Gordon},
  \citenamefont {Kouwenhoven}, \citenamefont {Marcus}, \citenamefont {Hirose},
  \citenamefont {Wingreen},\ and\ \citenamefont {Umansky}}]{Cronenwett2002}%
  \BibitemOpen
  \bibfield  {author} {\bibinfo {author} {\bibfnamefont {S.~M.}\ \bibnamefont
  {Cronenwett}}, \bibinfo {author} {\bibfnamefont {H.~J.}\ \bibnamefont
  {Lynch}}, \bibinfo {author} {\bibfnamefont {D.}~\bibnamefont
  {Goldhaber-Gordon}}, \bibinfo {author} {\bibfnamefont {L.~P.}\ \bibnamefont
  {Kouwenhoven}}, \bibinfo {author} {\bibfnamefont {C.~M.}\ \bibnamefont
  {Marcus}}, \bibinfo {author} {\bibfnamefont {K.}~\bibnamefont {Hirose}},
  \bibinfo {author} {\bibfnamefont {N.~S.}\ \bibnamefont {Wingreen}}, \ and\
  \bibinfo {author} {\bibfnamefont {V.}~\bibnamefont {Umansky}},\ }\href
  {http://link.aps.org/doi/10.1103/PhysRevLett.88.226805} {\bibfield  {journal}
  {\bibinfo  {journal} {Phys. Rev. Lett.}\ }\textbf {\bibinfo {volume} {88}},\
  \bibinfo {pages} {226805} (\bibinfo {year} {2002})}\BibitemShut {NoStop}%
\bibitem [{\citenamefont {Sfigakis}\ \emph {et~al.}(2008)\citenamefont
  {Sfigakis}, \citenamefont {Ford}, \citenamefont {Pepper}, \citenamefont
  {Kataoka}, \citenamefont {Ritchie},\ and\ \citenamefont
  {Simmons}}]{Sfigakis2008}%
  \BibitemOpen
  \bibfield  {author} {\bibinfo {author} {\bibfnamefont {F.}~\bibnamefont
  {Sfigakis}}, \bibinfo {author} {\bibfnamefont {C.~J.~B.}\ \bibnamefont
  {Ford}}, \bibinfo {author} {\bibfnamefont {M.}~\bibnamefont {Pepper}},
  \bibinfo {author} {\bibfnamefont {M.}~\bibnamefont {Kataoka}}, \bibinfo
  {author} {\bibfnamefont {D.~A.}\ \bibnamefont {Ritchie}}, \ and\ \bibinfo
  {author} {\bibfnamefont {M.~Y.}\ \bibnamefont {Simmons}},\ }\href
  {http://link.aps.org/doi/10.1103/PhysRevLett.100.026807} {\bibfield
  {journal} {\bibinfo  {journal} {Phys. Rev. Lett.}\ }\textbf {\bibinfo
  {volume} {100}},\ \bibinfo {pages} {026807} (\bibinfo {year}
  {2008})}\BibitemShut {NoStop}%
\bibitem [{\citenamefont {Wu}\ \emph {et~al.}(2012)\citenamefont {Wu},
  \citenamefont {Li}, \citenamefont {Zhang},\ and\ \citenamefont
  {Chang}}]{Wu2012}%
  \BibitemOpen
  \bibfield  {author} {\bibinfo {author} {\bibfnamefont {P.~M.}\ \bibnamefont
  {Wu}}, \bibinfo {author} {\bibfnamefont {P.}~\bibnamefont {Li}}, \bibinfo
  {author} {\bibfnamefont {H.}~\bibnamefont {Zhang}}, \ and\ \bibinfo {author}
  {\bibfnamefont {A.~M.}\ \bibnamefont {Chang}},\ }\href
  {http://link.aps.org/doi/10.1103/PhysRevB.85.085305} {\bibfield  {journal}
  {\bibinfo  {journal} {Phys. Rev. B}\ }\textbf {\bibinfo {volume} {85}},\
  \bibinfo {pages} {085305} (\bibinfo {year} {2012})}\BibitemShut {NoStop}%
\end{thebibliography}

%

\end{document}